\documentclass[referee]{aa}    		
\usepackage{graphicx}

\begin{document}

\title{Excess Hard X-ray Emission from the Obscured Low Luminosity AGN
In the Nearby Galaxy M~51 (NGC 5194)}

\author{
Yasushi Fukazawa\inst{1,3}
\and Naoko Iyomoto\inst{2,3}
\and Aya Kubota\inst{3}
\and Yukari Matsumoto\inst{3}
\and Kazuo Makishima\inst{3}\\
}

\offprints{Y. Fukazawa, \email{fukazawa@hirax6.hepl.hiroshima-u.ac.jp}}
\date{Received 29 January 2001 / Accepted 27 April 2001}

\institute{Department of Physical Sciences, Graduate School of Science,
Hiroshima University, 1-3-1 Kagamiyama, Higashi-Hiroshima, Hiroshima
739-8526
email:fukazawa@hirax6.hepl.hiroshima-u.ac.jp
\and Institute of Space and Astronautical Science, 3-1-1 Yoshinodai, Sagamihara, Kanagawa 229--8510
\and Department of Physics, Graduate School of Science, University of
Tokyo, 7-3-1 Hongo, Bunkyo-ku, Tokyo 113-0033
}

\abstract
{
We observed the nearby galaxy M~51 (NGC 5194) with BeppoSAX. 
The X-ray properties of the nucleus below 10
keV are almost the same as the ASCA results regarding 
the hard component and the neutral Fe K$\alpha$ line, but 
the intensity is about half of the ASCA 1993 data.
Beyond this, in the BeppoSAX PDS data, we detected a bright 
hard X-ray emission component which dominates above 10 keV.  
The 10 -- 100 keV flux and luminosity of this component are 
respectively $2\times10^{-11}$ erg s$^{-1}$ cm$^{-2}$ and 
$2\times10^{41}$ erg s$^{-1}$.  These are about 10 times
higher than the extrapolation from the soft X-ray band, and similar to the
flux observed with Ginga, which found a bright power law component in
2 -- 20 keV band.  
Considering other wavelength properties and the X-ray luminosity, 
together with strong neutral Fe K line, the hard X-ray emission 
most likely arises from a low luminosity active nucleus, which is 
obscured with a column density of $\sim10^{24}$cm$^{-2}$.
This suggests that hidden low luminosity AGNs may well be present 
in other nearby galaxies.  We interpret the discrepancy between Ginga 
and other X-ray satellites to be due to a large variability of 
absorption column density toward the line of sight over several years, 
suggesting that the Compton thick absorption material may be present 
on a spatial scale of a parsec.  Apart from the nucleus, 
several ultra-luminous off-nuclear X-ray sources detected in 
M~51 exhibit long-term time variability, suggesting the state 
transition similar to that observed in Galactic black hole candidates.
}


\maketitle

\section{Introduction}

M~51 is a face-on spiral galaxy, also known as the Whirlpool galaxy;  it is 
located at a distance of 9.6 Mpc (Sandage \& Tammann \cite{sandage}).
Optical studies of emission lines classified
it as a LINER (Low Ionization Nuclear Emission Region) 
or a Seyfert 2 galaxy (Stauffer \cite{stauffer}; Filippenko \&
Sargent \cite{filippenko}; Ho et al. \cite{ho97b}).  Furthermore, 
Ho et al. (\cite{ho97c}) suggested the presence of a broad H$\alpha$
line.  Kohno et al. (\cite{kohno}) found a nuclear molecular disk, 
and constrained the dynamical mass within 70 pc of the nucleus to be 
(4 -- 7)$\times10^6 M_{\odot}$.  This indicates that M~51 hosts as 
massive a black hole as many bright AGNs.

Observations in other wavelengths also strongly suggest the presence of 
an AGN in M~51, although
previous X-ray observations of M~51 gave puzzling results.
Einstein (Palumbo et al. \cite{palumbo}) and ROSAT (Marston et
al. \cite{marston}; Ehle et al. \cite{ehle}) observations 
constrained the soft X-ray
luminosity of the M~51 nucleus to be $L_{\rm X}<5\times10^{39}$erg s$^{-1}$. 
Ginga scanning observations in 1988 detected bright hard X-ray 
emission with $L_{\rm X}=1.2\times10^{41}$erg/s (2 -- 20 keV) with a photon 
index of $1.43\pm0.08$ (Makishima et al. \cite{makishima90}) and 
an intrinsic absorption of $<7\times10^{21}$cm$^{-2}$.
Such hard nuclear X-ray emission is often considered to be 
evidence of Low Luminosity AGN (LLAGN), 
but it should have been detected with Einstein and ROSAT if the flux did 
not change significantly.
After the Ginga observations, ASCA observed M~51 in the hard X-ray band 
(2 -- 10 keV) in
1993, and did not detect such a bright hard component (Terashima et
al. \cite{terashima98a}).  
Instead, a faint hard X-ray continuum with a neutral Fe K line was
detected, whose flux was an order of magnitude lower than that measured by 
Ginga.  One possibility is that the difference between these observations 
is due to time variability with a large amplitude.
Such a strong time variability is not often observed for low luminosity
AGNs, 
and therefore is quite interesting if it is correct.
The other possibility is a strong change of the intrinsic absorption.
The key phenomenon that can help in resolving 
this mystery is the presence of a neutral Fe K line emission
detected by ASCA (Terashima et al. \cite{terashima98a}), 
which indicates that during the ASCA observations, the nucleus of 
M~51 was heavily obscured by Compton thick material below 10 keV, 
as found in some Seyfert 2 galaxies (e.g. Risaliti et al. \cite{risaliti99}).
This idea can be confirmed by X-ray observations at energies above 10 keV.

Detection of an obscured hard X-ray emission from M~51 would suggest that
many galaxies may harbor previously unknown hidden LLAGNs emitting 
below 10 keV.  Moreover, such a detection can be valuable 
for a study of the physical structure of Compton thick absorbing material 
around AGNs.  To further investigate such a possibility, 
we performed an X-ray observation of M~51 with BeppoSAX, sensitive to 
hard X-ray emission (Boella et al. \cite{boella97b}).

\section{Observations and Data Reduction}

We observed M~51 with BeppoSAX on Jan. 18--20, 2000.
The three Narrow Field Instruments, LECS (Parmar et al. \cite{parmar}),
MECS2+3 (Boella et al. \cite{boella97b}), and PDS 
(Frontera et al. \cite{frontera}) collected the data during the observation.
We obtained the net exposures of 37 ks, 97 ks, and 45 ks, for LECS, MECS, and
PDS, respectively.
We utilized the cleaned and linearized data supplied by the BeppoSAX
Science Data Center.
The light curves, spectra, and images were reduced with the standard
software, such as XSELECT and FTOOLS-4.2.
Throughout this paper, the data of two MECS instruments were summed
together.
The cross-normalization constant between MECS and PDS is set to be 0.85
in this paper (Fiore et al. \cite{fiore99}).

In addition, we utilized ASCA (Tanaka et al. \cite{tanaka}) archival 
data in order to 
investigate the time variability of the M~51 nucleus.
ASCA observed M~51 twice, on May 11--13, 1993 and May 5--7, 1994.
The results of ASCA 1993 observation were reported by Terashima 
et al. (\cite{terashima98a}).  Ptak et al. (\cite{ptak}) also 
performed the analysis of both 1993 and 1994 data, but they did 
not consider the contamination by the neighboring off-nuclear
bright X-ray source.  The screened data were obtained from the 
ASCA archival data center.  In the data reduction, we imposed 
the selection criterion on the GIS data requiring that the 
rigidity is $>6$ GeV c$^{-1}$ (Ohashi et al. \cite{ohashi};
Makishima et al. \cite{makishima96}).  The net GIS exposure 
is 35--38 ksec for both observations, but it was only 12 ksec for
the SIS data in 1994 due to problems of the onboard data processing.

\section{Image Analysis}

Before spectral analysis can be performed on the nuclear component, 
it is necessary to investigate X-ray images because M~51
is known to contain several non-nuclear X-ray sources.
In Fig. 1, we show the X-ray images taken with the BeppoSAX
MECS.  The MECS image in the total band is 
overlaid on the optical image taken from SkyVIEW from NASA/GSFC.
Three X-ray sources are clearly visible:  the nucleus of
M~51, a source in the companion galaxy NGC 5195 (source A), and a source 
to the east of the nucleus (source B).  
The nucleus is particularly bright in the 
5.7 -- 7.0 keV band, where a neutral Fe K line exists.
Recently, Terashima et al. (\cite{terashima01}) confirmed that this
source is resolved as a nuclear point source with Chandra.
The source B is not the supernova SN1994I, since its position is
different and the flux observed in ROSAT is quite low 
(Immler et al. \cite{immler}).

In Fig. 2, we show ASCA GIS images of M~51 taken in 1993 and 1994.  
The nucleus and the source A are always visible, in the ASCA 1993 
and BeppoSAX 2000 observations.  The source B is not seen in the 
ASCA GIS images, indicating the time variability of this source.
On the other hand, the north-east X-ray source (source C) 
that was visible in the ASCA 1994
observation is not seen in the ASCA 1993 and BeppoSAX image.

Sources B and C are thought to correspond to 
the ROSAT PSPC point sources R3 and R1 in Marston et
al. (\cite{marston}), respectively.
All the detected sources have X-ray luminosities in the range of 
$\sim10^{39-40}$ erg s$^{-1}$.

\section{Spectral Analysis}

\subsection{BeppoSAX Spectra of the Nucleus}

In order to investigate the spectrum of the 
nucleus, we extracted data from the imaging instruments onboard BeppoSAX 
corresponding to a radius of 2$'$ 
so that neighboring X-ray sources
do not contaminate the spectra of the nuclear source.  
The background spectra were prepared using data from 
the same detector region, utilizing
the standard background data set (Nov. 1998 for MECS, Dec. 1999 for LECS), 
and we subtracted them from the on-source spectra.  In the spectral fitting, 
we used the standard response matrices released in September 1997.  

In Fig. 3, we show the X-ray spectra of LECS and MECS.
The data cannot be fit by a simple power law;  
they reveal both Fe-L and Fe K line features, 
but the Raymond-Smith thermal model 
(Raymond \& Smith \cite{raymond}) with the Galactic absorption of
$1.5\times10^{20}$cm$^{-2}$ (Stark et al. \cite{stark}) 
does not fit the data either:  the reduced chi-square is $\sim2$ for
both models.
These spectral features are similar to the ASCA 1993 spectra.
Thus, we fit the spectra with a model including Raymond-Smith thermal plasma 
plus a power law, and including an additional narrow line.
The metal abundances are fixed to $0.2 \times$ Solar with the Solar 
metal abundance ratios (Anders \& Grevesse \cite{anders}).
Such a model was previously applied to the ASCA 1993 spectra (Terashima et al. 
\cite{terashima98a}).  The fit is acceptable as shown in Fig. 3 and Table 1.
The power law photon index of $1.59\pm0.23$ is not well constrained, 
but it is consistent with the ASCA results.  The Fe K line center 
energy is $6.47\pm0.14$ keV, indicating that it is due to fluorescence 
from the neutral iron, in agreement with the ASCA results.
The equivalent width of the Fe K line is $0.9_{-0.5}^{+0.4}$ keV, 
which is within the error range of the 1993 ASCA results.  
The temperature of the Raymond-Smith plasma is $0.51\pm0.16$ keV.  
The flux and luminosity of the power law 
component are $3.4_{-0.3}^{+0.4}\times10^{-13}$ erg s$^{-1}$
cm$^{-2}$ and $3.6_{-0.3}^{+0.4}\times10^{39}$ erg s$^{-1}$,
respectively in 2 -- 10 keV band, about 30\% of 1993 ASCA results.  
The strength of Fe K line is also smaller than that of 1993 ASCA 
value of $(4-14)\times10^{-6}$ c s$^{-1}$ cm$^{-2}$.  

We conclude that the spectral properties below 10 keV measured in 
the BeppoSAX data are almost the same as the ASCA results except 
for the flux. To extend the bandpass beyond that of the imaging 
instruments (10 keV), we plot a PDS spectrum together with 
the LECS and MECS spectra in Fig. 4a.  The solid line represents 
the best-fit model obtained by the LECS and MECS spectral analysis.  
It can be seen that the PDS flux is higher by a factor of
$\sim10$ compared with the extrapolation of the best-fit model of 
LECS and MECS, and significant photon flux is detected up to 50 keV.  
This spectral feature is quite similar to that seen in Compton-thick 
Seyfert 2 galaxies such as NGC 4945 (Done et al. \cite{done}), Circinus 
Galaxy (Matt et al. \cite{matt99}), and Mrk 3 (Cappi et al. \cite{cappi})

We add an absorbed power law component to represent this hard
X-ray emission, and fit the LECS + MECS + PDS spectra simultaneously.
The best-fit results are shown in Table 1 and Fig. 4b.
The wide-band X-ray spectra require a hard X-ray component strongly absorbed
by optically thick material with a column density $\sim10^{24}$ cm$^{-2}$.
The best-fit parameters other than this newly included component do not
change significantly.  Since the photon index and absorption column 
density for the hard X-ray component are not well constrained separately
(as they are correlated in spectral fits), to estimate the absorbing column 
we fixed the photon index at 1.9, typical for Seyfert 1 galaxies 
without a reflection component.  Then, the absorption column density 
becomes $5.6_{-1.6}^{+4.0}\times10^{24}$ cm$^{-2}$, and the X-ray flux 
and luminosity are $1.9\times10^{-11}$ erg s$^{-1}$ cm$^{-2}$ and
$2.0\times10^{41}$ erg s$^{-1}$, respectively in 10 -- 100 keV band.
The absorption -- corrected X-ray flux is $2.2\times10^{-11}$ 
erg s$^{-1}$ cm$^{-2}$ in 2 -- 20 keV band, similar to the Ginga results, and
$1.5\times10^{-11}$ erg s$^{-1}$ cm$^{-2}$ in 2 -- 10 keV which should be 
easily detected with ASCA if no absorption was present.  
When we fit only the PDS spectrum with an absorbed power law model by fixing
the power law photon index to be 1.9, 
we obtained only the upper limit of $N_{\rm H}<5.4\times10^{24}$cm$^{-2}$.
This is most likely because the model adopted by us, with a luminous but 
heavily absorbed hard component plus a soft component, has an apparent absorption feature 
located in the energy band between MECS and PDS.
To investigate a possible presence of a Compton reflection component, 
we include it by fitting the data to the spectral model ``pexrav''
as implemented in XSPEC;  with this, the reflection fraction becomes $<$18\%.

\begin{table*}
\scriptsize
\caption[]{Results of joint spectral fittings of M~51 nucleus with various models.}
\begin{center}
\begin{tabular}{cccccccccc}
\hline
\hline
$kT$ & norm & $\alpha_{\rm ph}^s$ & norm$^s$ & $E_{\rm Fe}^g$ & norm$^g$ & $N_{\rm H}^h$ & $\alpha_{\rm ph}^h$ & norm$^h$ & $\chi^2$/dof \\
(keV) & ($10^{10}$cm$^{-5}$) & & $10^{-3}$ & (keV) 
 & $10^{-6}$ & $10^{24}$cm$^{-2}$ & & & \\
\hline
\multicolumn{4}{l}{LECS + MECS} & \multicolumn{6}{l}{ABS$^s$*PO$^h$}\\
& & $1.86_{-0.18}^{+0.20}$ & $11.1_{-0.22}^{+0.29}$ & & & & & & 2.19 \\
\multicolumn{4}{l}{LECS + MECS} & \multicolumn{6}{l}{ABS$^s$*(RS+PO$^s$+GA)}\\
$0.51\pm0.16$ & $4.4_{-1.4}^{+2.3}$ & $1.59\pm0.24$ & $7.1_{-2.1}^{+2.6}$ & $6.47\pm0.14$ & $3.4\pm1.5$ & & & & 1.18 \\
\hline
\multicolumn{4}{l}{LECS + MECS + PDS} & \multicolumn{6}{l}{ABS$^s$*(RS+PO$^s$+GA+ABS$^h$*PO$^h$)}\\
$0.52_{-0.17}^{+0.15}$ & $4.4_{-1.4}^{+2.3}$ & $1.59\pm0.23$ & $7.0_{-2.0}^{+2.7}$ & $6.47\pm0.14$ & $3.4\pm1.5$ & 10$_{-5}^{+25}$ & $3.3_{-1.0}^{+3.0}$ & --- & 1.14 \\
\multicolumn{4}{l}{LECS + MECS + PDS} & \multicolumn{6}{l}{ABS$^s$*(RS+PO$^s$+GA+ABS$^h$*PO$^h$)}\\
$0.52_{-0.17}^{+0.15}$ & $4.40_{-1.4}^{+2.3}$ & $1.60\pm0.23$ & $7.2_{-1.9}^{+2.5}$ & $6.47\pm0.14$ & $3.4\pm1.5$ & 5.8$_{-1.8}^{+3.8}$ & 1.9 fix & $4.9_{-1.5}^{+1.7}$ & 1.21 \\
\hline
\multicolumn{10}{p{15cm}}{ABS$^s$ and ABS$^h$: Photoelectric absorption
 of the Galactic and intrinsic material, RS:
 Raymond-Smith thermal model, PO$^s$ and PO$^h$: Power law model, GA:
 Gaussian line.}\\
\multicolumn{10}{p{15cm}}{$g$: Parameters of a gaussian model.}\\
\multicolumn{10}{p{15cm}}{The unit of normalizations (norm) of power law and
 gaussian is c s$^{-1}$ cm$^{-2}$ keV$^{-1}$ at 1 keV and c s$^{-1}$ cm$^{-2}$, respectively.}\\
\end{tabular}
\end{center}
\normalsize
\end{table*}

\subsection{ASCA spectra of the Nucleus}

Since the result of the 1993 ASCA observation are reported in Terashima et
al. (\cite{terashima98a}), we here show the result of the 1994 observation.
Since the exposure time of the SIS is short, we created only the GIS 
spectrum.  The source C is located at $\sim2'$ from the nucleus.
Thus we created spectra using counts detected only within 1.5$'$ 
of the nucleus, where the source C may contaminate the nuclear 
spectrum by $\sim$20\%.  The spectral model used for the fits of 
the ASCA data is the same as that for LECS and MECS spectra:  
Raymond-Smith plasma, plus power law, and one emission line. 
The normalizations of thermal and power law components are left free, but 
the other model parameters are fixed to the values measured in the 1993 
observation, due to the poor quality of 1994 data.

The spectrum and best-fit model are shown in Fig. 5, which tells us
that the overall property of the spectrum apart from the normalization 
is almost the same as that of ASCA 1993 data and BeppoSAX data.  
The X-ray flux of the power law component is $\sim3\times10^{-13}$ 
erg s$^{-1}$ cm$^{-2}$ in 2 -- 10 keV band, about 30\% of that 
measured in the 1993 data and almost the same as BeppoSAX data.
The details of the Fe K line are ambiguous due to the small extraction area.  .
In order to constrain the Fe K line strength better, we extended
the integration radius to 3.75$'$.  Taking into account that about 
45\% of photons in this spectrum come from the source C, we 
obtain an upper limit on the Fe K line strength of $2.7\times10^{-6}$ 
photons s$^{-1}$ cm$^{-2}$, significantly smaller than that measured in the 
1993 ASCA data.

\subsection{Time Variability of the Nucleus}

We investigate the short time variability properties of the hard 
component of the nucleus. We make light curves in the 
3 -- 10 keV band (MECS) and 10 -- 100 keV band (PDS).  The MECS light 
curve shows a counting rate from the region within 2$'$ of the nucleus
without background subtraction, since the count rate of the background is only 
25\%.  The PDS light curve is created after background subtraction.
We do not find any significant short-term time variability in either light 
curve.  

In Fig. 6, we show the long-term time variability of the X-ray flux of the
power law component in the 2 -- 10 keV band and absorption-corrected 2 -- 20 
keV band.  The X-ray flux in 2 -- 10 keV band represents the scattered 
component of the nuclear emission, and in 1994, it drops to 30\% of the 
1993 value.  Its flux in the 2000 BeppoSAX data is almost the same as that 
of 1994 ASCA observation.  The absorption-corrected flux in 2 -- 20 keV of 
Ginga and BeppoSAX PDS corresponds to the direct nuclear emission.
Nominally the 2000 BeppoSAX flux is twice the 1988 Ginga value, 
but the errors are large, and these fluxes are consistent with 
the values being the same.  The Fe line equivalent width is constant within 
the error range as described in \S4.2.

\subsection{Off-nuclear X-ray Sources}

M~51 includes several off-nuclear point-like sources.
The typical X-ray luminosity of the brightest X-ray sources is
$10^{39-40}$ erg s$^{-1}$, much higher than the Eddington limit for 
accreting neutron stars.  In the ASCA and BeppoSAX observations, 
three off-nuclear sources were visible (Figs. 1 and 2).  The source 
A is associated with NGC 5195, and its flux appears constant.  The 
other two sources (B and C) are located in the spiral arms.  They 
exhibit strong time variability, and are visible when they become 
as bright as the nucleus.

The source B shows a large time variability in 2000 as
shown in Fig. 7, where the count rate in the second half of the 
observation decreases to $\sim$50\% of that measured in the first half.  
This is the third example of short-term time variability of the off-nuclear 
ultra-luminous X-ray sources, 
after those in IC342 (Okada et al. \cite{okada};  
Kubota et al. \cite{kubota}) and M82 (Matsumoto et al. \cite{matsumoto}).
The spectra of the sources A and B are determined from the counts collected 
within 2$'$ of their position.  For the source A, we prepared the spectra 
using the ASCA 1993 data and the BeppoSAX data.  We do not use the ASCA 1994 
data because the GIS instrument cannot resolve the source A clearly 
(see Fig. 2).  The source B is only visible in the BeppoSAX data, 
and its count rate varies significantly.  We thus analyzed two MECS spectra; 
one is integrated in the bright phase (0 -- 110,000 sec in Fig. 7), 
and the other is in the faint phase (110,000 -- 190,000 sec in Fig. 7).

The X-ray spectra of these ultra-luminous off-nuclear X-ray
sources are known to be generally described by multi-color disk 
black body (MCD: Mitsuda et al. \cite{mitsuda};  Makishima et al. 
\cite{makishima86}; Mizuno et al. \cite{mizuno99}; Makishima et al. 
\cite{makishima00}; Kotoku et al. \cite{kotoku}), and thus we try 
to fit the spectra by a power law model and MCD model.  The spectrum 
of source A is almost the same between the 1993 ASCA and BeppoSAX 
observations, and thus we fitted both data sets simultaneously.
We performed spectral fittings with the energy band above 2 keV, to
avoid the uncertainty of the diffuse thermal emission due to the different
angular resolution among instruments.
Instead, we include the Raymond-Smith thermal model to represent the 
diffuse emission.  Its temperature cannot be well determined, and thus
we fixed the temperature of Raymond-Smith model to 0.39 and 0.35 keV 
for source A and B, respectively, based on the spectral fitting of the 
ASCA SIS data around each source.  Since we are interested in the 
hard component rather than soft thermal emission, we performed the 
spectral fitting above 2 keV.  The results are summarized in Table 2 
and Fig. 8.

The limited photon statistics does not allow us to distinguish between these 
two models for both sources.  
The inner disk temperature of MCD model for the source A, 1.5 -- 2.2 keV, is
typical for ultra-luminous off-nuclear X-ray sources in other galaxies.
The luminosity of the MCD component in the 2 -- 10 keV band is 
$3.2\times10^{39}$ erg s$^{-1}$ for source A.  
For the source B, the spectrum in the high flux period seems to be
harder than that in the low flux period; the power law photon index or 
the temperature of the MCD model is significantly different.
We expect a constant radius of the MCD model despite different
temperatures, but that cannot be tested due to the poor data
quality.
The luminosities of the MCD component in the 2 -- 10 keV band of the source B are  
$3.8\times10^{39}$ erg s$^{-1}$ and $1.2\times10^{39}$ erg s$^{-1}$ for
the bright and faint state, respectively.  

In order to study the long-term time variability of these sources, we
utilized the archival ROSAT HRI data from HEASARC (NASA/GSFC High 
Energy Astrophysics Science Archive Research Center) and the paper of
Ehle et al. (\cite{ehle}).  We cannot resolve the source A clearly 
in the HRI images, and thus the source A may be an assembly of discrete 
sources in NGC 5195.  The constant flux is consistent with this scenario.  
The flux histories for the sources B and C are summarized in Table 3.
These are typically in the range of (0 -- 3)$\times10^{-13}$ erg cm$^{-2}$ 
s$^{-1}$, at a level comparable to the ASCA and BeppoSAX measurements.  
The source B varies by a factor of 2, while the variability of the source 
C is quite strong, by a factor of $>$5.

\begin{table*}
\caption[]{Results of spectral fittings of source A and B with various models above 2 keV.}
\begin{center}
\begin{tabular}{ccccccc}
\hline
\hline
$kT^r$ & norm$^r$ & $\alpha_{\rm ph}^p$ & norm$^p$ & $kT_{\rm in}^d$ & norm$^d$ & $\chi^2$/dof \\
(keV) & ($10^{11}$cm$^{-5}$) & &  & (keV) & $10^{-3}$ &  \\
\hline
\multicolumn{7}{l}{Source A: GIS + SIS + LECS + MECS} \\
\multicolumn{7}{l}{Power law Model} \\
0.39 Fix & $0.4_{-0.4}^{+4.6}$ & $2.19_{-0.35}^{+0.25}$ & $1.77_{-0.73}^{+0.64}$ & & & 1.01 \\
\multicolumn{7}{l}{MCD Model} \\
0.39 Fix & $6.0_{-3.6}^{+3.5}$ & & & $1.80_{-0.31}^{+0.44}$ & $2.0_{-1.1}^{2.5}$ & 1.00 \\
\hline
\multicolumn{7}{l}{Source B bright phase: LECS + MECS} \\
\multicolumn{7}{l}{Power law Model} \\
0.35 Fix & $0.21_{-0.19}^{+4.18}$ & $2.10\pm0.17$ & $1.78_{-0.37}^{+0.42}$ & & & 1.04 \\
\multicolumn{7}{l}{MCD Model} \\
0.35 Fix & $0.53_{-0.16}^{+0.18}$ & & & $1.50_{-0.18}^{+0.21}$ & $5.3_{-2.1}^{+3.0}$ & 1.12 \\
\hline
\multicolumn{7}{l}{Source B faint phase: LECS + MECS} \\
\multicolumn{7}{l}{Power law Model} \\
0.35 Fix & $<0.33$ & $3.09_{-0.41}^{+0.31}$ & $2.36_{-0.86}^{+0.65}$ & & & 0.39 \\
\multicolumn{7}{l}{MCD Model} \\
0.35 Fix & $0.40_{-0.32}^{+0.30}$ & & & $0.70_{-0.19}^{+0.29}$ & $73_{-60}^{+333}$ & 0.51 \\
\hline
\multicolumn{7}{p{13cm}}{\small $r$: parameters of Raymond-Smith model}\\
\multicolumn{7}{p{13cm}}{\small $p$: parameters of Power law model}\\
\multicolumn{7}{p{13cm}}{\small $d$: parameters of MCD model}\\
\multicolumn{7}{p{13cm}}{The unit of norm$^p$ is $10^{-3}$c s$^{-1}$
 cm$^{-2}$ keV$^{-1}$ at 1 keV.}\\
\multicolumn{7}{p{13cm}}{The norm$^d$ represents
 $((Rin/\mbox{1km})/(D/\mbox{10kpc}))^2cos\theta$ where $Rin$ and
 $\theta$ are an inner radius and an inclination angle of the disk,
 respectively, and $D$ is a distance to the source.}
\end{tabular}
\end{center}
\end{table*}

\begin{table*}
\caption[]{ROSAT HRI Flux of the Source B and C}
\begin{center}
\begin{tabular}{cccccc}
\hline
\hline
Time & 1991.12.07 & 1992.01.08    & 1994.05.22 -- 23 & 1994.06.18 -- 24 &
 1997.12.26 -- 30 \\
     &            & 1992.05.22 -- 25 & & & \\
     &            & 1992.06.05    & & & \\
\hline
exposure & 16097s & 10411s & 9392s & 36343s & 8104s \\
B & 1.5 & 1.2    & 1.8 & 2.1 & 0.9 \\
C & 2.2 & $<$1.0 & 2.4 & 1.7 & $<$0.4 \\
\hline
\multicolumn{6}{p{13cm}}{The unit is $10^{-13}$ erg s$^{-1}$ cm$^{-2}$. The
 conversion from counts to source flux density is  $4.08\times10^{-11}$
 erg cm$^{-2}$ count$^{-1}$, from Ehle et al. (\cite{ehle}). The data in 1992 are quoted from Ehle et al. (\cite{ehle}).}
\end{tabular}
\end{center}
\end{table*}

\section{Discussion}

\subsection{Bright Hard X-ray Emission}

In our observations of M~51 with BeppoSAX, we detected in the PDS data 
bright hard X-ray emission above 10 keV, which we interpret as associated 
with an active nucleus present in the galaxy.  This inference is 
supported by the fact that in the MECS data we detect a neutral Fe K line 
emission from the central point-like source, and such emission is 
consistent with the ASCA observations.  The flux above 10 keV is about 
10 times higher than the extrapolation of the nuclear flux detected 
by the MECS below 10 keV.  Considering these spectral features, we 
infer that M~51 harbors a heavily obscured Seyfert 2 type nucleus, 
with the column density $10^{24}$ cm$^{-2}$ 

The X-ray luminosity of this hard component is $2.0\times10^{41}$
erg s$^{-1}$ in the 10 -- 100 keV band, and thus M~51 is probably the first
low-luminosity Compton thick Seyfert 2 AGN.  The existence of AGN 
with X-ray luminosity of $\sim10^{41}$ erg s$^{-1}$ is reasonable, 
according to the prediction from the correlation of H$\alpha$ and 
X-ray luminosity in Seyfert 1 galaxies (Ward et al. \cite{ward}), 
and the correlation of [O$_{\rm III}$] and X-ray luminosity of 
Seyfert galaxies (Mulchaey et al. \cite{mulchaey}), based on the 
M~51 value of $L_{\rm OIII}\sim L_{{\rm H}\alpha}\sim10^{39}$ erg s$^{-1}$.
Moreover, the presence of an obscured nucleus in M~51 is suggested by 
a dense molecular disk with $>3\times10^{23}$cm$^{-2}$ detected around
the nucleus by HCN and CO observations (Kohno et al. \cite{kohno}).

Nevertheless, we cannot completely rule out the possibility that the 
PDS spectrum is contaminated by bright hard sources, because the
absorption is ambiguous if one considers the PDS spectrum alone.  
However, there are several observational results which support the above
conclusion.  The Ginga observation -- using the scanning mode, and thus 
allowing good positional accuracy -- indicated that the bright hard X-ray 
emission indeed comes from the M~51 nucleus (Makishima et al. 
\cite{makishima90}).  Since the hard X-ray flux is similar between 
the Ginga LAC and the BeppoSAX PDS, the hard emission detected from the 
PDS is also likely to come from M~51.  In addition, there are no other 
bright X-ray point sources with a flux $>10^{-12}$ erg s$^{-1}$ 
cm$^{-2}$ (0.1 -- 2 keV) within 2 degrees of the M~51 nucleus in the 
ROSAT PSPC image.  Consequently, we conclude that the hard X-ray 
emission detected in the PDS is associated with the nucleus of M~51.  
Then, M~51 is found to be one of the brightest low luminosity AGNs 
in the sky above the 10 keV band, together with M~81 and NGC 4258
(Pellegrini et al. \cite{pellegrini}; Fiore et al. \cite{fiore}).
It is interesting that these two objects are not Compton thick, unlike M51.

There still remains a difference between the Ginga and other satellite
observations; the X-ray flux below 10 keV obtained with Ginga was higher 
than that measured by other instruments by a factor of more than 20,
and the X-ray spectrum obtained with Ginga exhibited no absorption features.  
One possibility is the change of the column density of the absorbing 
material in the line of sight; the absorption happened to be weak in the 
Ginga observation.  Such a case is in fact observed for the famous Seyfert 
galaxy NGC 4051 (Uttley et al. \cite{uttley}), and for NGC 1365 (Risaliti et 
al. \cite{risaliti00}) with BeppoSAX, and thus such a behavior is not unique 
to M~51.  Another scenario would have a large change in the soft ($< 10$ keV) 
flux, by a factor of more than 20.  However, the X-ray flux above 10 keV band 
is almost the same between the Ginga and BeppoSAX observation epochs, and in 
addition agrees with the prediction from $L_{\rm OIII}$ and 
$L_{{\rm H}\alpha}$, suggesting that the former explanation is more likely.

\subsection{A Neutral Fe K line and Reflected Continuum}

BeppoSAX detected neutral Fe K lines from the M~51 nucleus, confirming the 
previous ASCA detection (Terashima et al. \cite{terashima98a}).  This 
line is likely to arise via fluorescence from neutral material.
The equivalent width of the Fe K line, $0.9_{-0.5}^{+0.4}$ keV is consistent 
with the ASCA 1993 result (Terashima et al. \cite{terashima98a}).
The equivalent width of the Fe K line can be explained as arising via 
reflection by the reprocessing material covering $\sim2\pi$ around the nucleus
with $1 \times$ Solar metal abundances.  Other well-known examples of objects 
indicating such geometry are NGC 6552 (Fukazawa et al. \cite{fukazawa}) 
and Circinus Galaxy (Matt et al. \cite{matt96}).  

A natural association of the Fe K line would be with the dense, neutral 
material present around the nucleus;  such material is most likely 
responsible for the Compton reflection component, seen in many Seyfert 1 
galaxies.  We investigated how strong of such reflection component is in 
the BeppoSAX data, but obtained a small upper limit corresponding to a 
solid angle subtended by the reflector of 18\% of 2$\pi$.  If the material 
responsible for the Fe K line and the Compton reflection were to be the same, 
this value seems to be in contradiction with the Fe K equivalent width.  
A more likely scenario is that the Fe K line is produced in a Compton-thick 
region {\sl in transmission}.  In fact, the equivalent width of the Fe K 
line can be reproduced by such a situation with a column density of 
$\sim10^{24}$cm$^2$, covering factor of $2\pi$, and a Solar Fe abundance 
(Makishima \cite{makishima86a}).

\subsection{Time variability}

The bright hard component that was detected with Ginga and BeppoSAX, 
presumably due to direct nuclear emission, has almost constant flux 
during the BeppoSAX observation.  Such a trend of a weak time variability 
is similar to that seen in other LLAGNs (Ishisaki et al. \cite{ishisaki};  
Iyomoto et al. \cite{iyomoto96}, \cite{iyomoto00}; Iyomoto \cite{iyomoto99}; 
Reynolds et al. \cite{reynolds}), and is not consistent with the trend 
seen in Seyfert galaxies (Nandra et al. \cite{nandra}; Hayashida et al. 
\cite{hayashida}).  We note that a scenario involving an advection 
dominated accretion flow (ADAF) is consistent with such a behavior, 
but any more detailed discussion of the applicability of ADAFs to LLAGN is 
beyond the scope of this paper (but see, e.g., Lasota et al. \cite{lasota}).  

On the other hand, we detected time variability of two quantities:  
the weak hard X-ray emission below 10 keV, and the absorption column 
density for direct nuclear emission.  These are indicative of a change of 
conditions of the material surrounding the nucleus.  The former, 
which is thought to be a scattered or reflected nuclear component, 
exhibits a long time variability, by a factor of 2 -- 3 on a time scale 
of years, through ASCA and BeppoSAX observations (Fig. 7).  The strength 
of the Fe K line also varied simultaneously with the continuum by almost 
the same factor.  Regarding the latter, we measure the intrinsic 
absorption column density changes by a factor of $>100$ within several 
years between the Ginga and ASCA observations.  These indicate that the 
Compton-thick absorption material has a spatial scale 
(meaning the size and/or a distance from the nucleus) of a parsec.
Such a parsec -- scale torus material is also suggested by other 
observations (Greenhill et al. \cite{greenhill}; Gallimore et al. 
\cite{gallimore}; Siebenmorgen et al. \cite{siebenmorgen}; Matt 
\cite{matt00}).  Further variability studies of direct nuclear 
emission on shorter time scales at energies above 10 keV, such as 
with INTEGRAL and Astro-EII HXD, are necessary to characterize the 
dimensions of this component in more detail.  

\subsection{Obscured Low Luminosity AGNs}

ASCA survey showed that LINER galaxies
often exhibit hard X-ray emission  with X-ray luminosity of 
$(2-5)\times10^{40}$erg/s (2-10 keV) and power law spectra with 
photon indices of $\sim$1.7  
(Makishima et al. \cite{makishima94}; Ishisaki et al. \cite{ishisaki}; 
Iyomoto et al. \cite{iyomoto96}; \cite{iyomoto97}; \cite{iyomoto98};  
Terashima et al. \cite{terashima98a}; \cite{terashima98b}; \cite{terashima00}). 
However, there are LINERs which do not show a significant 
excess of hard X-ray emission below 10 keV.  Interestingly, ASCA results 
show that some LLAGNs also exhibit an intrinsic absorption 
of $\sim10^{23}$cm$^{-2}$ which is comparable to Compton-thin 
Seyfert 2 galaxies.  Furthermore, ASCA found two LLAGNs, M~51 and NGC 1365 
(Terashima et al. \cite{terashima98a}; Iyomoto et al. \cite{iyomoto97}), which 
exhibit a flat hard X-ray component with a strong neutral Fe K line, and thus
these are good candidates for the low luminosity Compton-thick Seyfert 2 
galaxies.  In this context, our BeppoSAX observations confirm 
M~51 as a Compton-thick LLAGN.
These results suggest that the distribution of the absorption column density 
of LLAGNs may be similar to that of Seyfert galaxies, 
and thus a significant fraction of LINERs may be also completely obscured 
below 10 keV, as is the case for Seyfert galaxies.
LLAGNs are thought to be quite numerous because LINERs are quite 
numerous (Keel \cite{keel}; Ho et al. \cite{ho97a}).  With this, 
future X-ray observations of LINER galaxies may reveal 
many LLAGNs in galaxies that are considered otherwise ``normal.''
Understanding the relationship between LINERs and LLAGN is likely 
to be an important step in understanding the link between 
evolution of normal and active galaxies.

\subsection{Ultra-Luminous Off-nuclear X-ray Sources}

M~51 contains several ultra-luminous off-nuclear X-ray sources found with
Einstein and ROSAT (Palumbo et al. \cite{palumbo}; Marston et
al. \cite{marston}; Ehle et al. \cite{ehle}).
ASCA and BeppoSAX observations resolved three of them, and all three 
exhibit time variability of a factor of $\sim3$.
The X-ray luminosity measured in those sources -- 
$10^{39-40}$ erg s$^{-1}$ -- cannot be explained by
neutron star binaries, as the Eddington limit for those is $2\times10^{38}$
erg s$^{-1}$.  The X-ray spectra for two of them can be fitted 
with power law or disk blackbody models.  This spectral shape is 
similar to that seen in other ultra-luminous off-nuclear X-ray 
sources in other galaxies (Mizuno et al. \cite{mizuno99};
Makishima et al. \cite{makishima00}; Mizuno \cite{mizuno00}).
These bright X-ray sources in nearby galaxies are postulated to be 
black hole binaries, where spectra (and in particular, the disk 
black body temperature and normalization) imply Kerr black holes with 
masses $\sim$100 $M_{\odot}$ (Mizuno et al. \cite{mizuno99}; Makishima et al. 
\cite{makishima00};  Mizuno \cite{mizuno00}).  We suggest that the 
bright, off-nuclear X-ray sources in M~51 detected with ASCA and BeppoSAX
are objects of the same type.  The strong variability of these sources 
is similar to that often observed in the Galactic black hole candidates.
Since their X-ray spectra were obtained only in their 
bright states, it would be quite interesting to know what are the X-ray
spectra in their faint states, and in particular, whether they 
exhibit hard power law spectra in their faint states, similar
to the Galactic black hole candidates.  So far, two ultra-luminous off 
nuclear X-ray sources in IC342 exhibited such a spectral transition 
(Kubota et al. \cite{kubota}), suggesting that these are black-hole binaries.
Since it contains relatively many ultra-luminous off-nuclear X-ray
sources, M~51 is an interesting target for repeated observations by Chandra
and XMM-Newton to study such sources.  

We are grateful to Prof. Greg Madejski for correction of the English in the 
manuscript.
The authors also thank the anonymous referee for careful reading and
helpful comments, and thank as well
operation and calibration teams of BeppoSAX and ASCA.

\begin{figure}
\begin{minipage}{8cm}
\centerline{\resizebox{8cm}{!}{\includegraphics{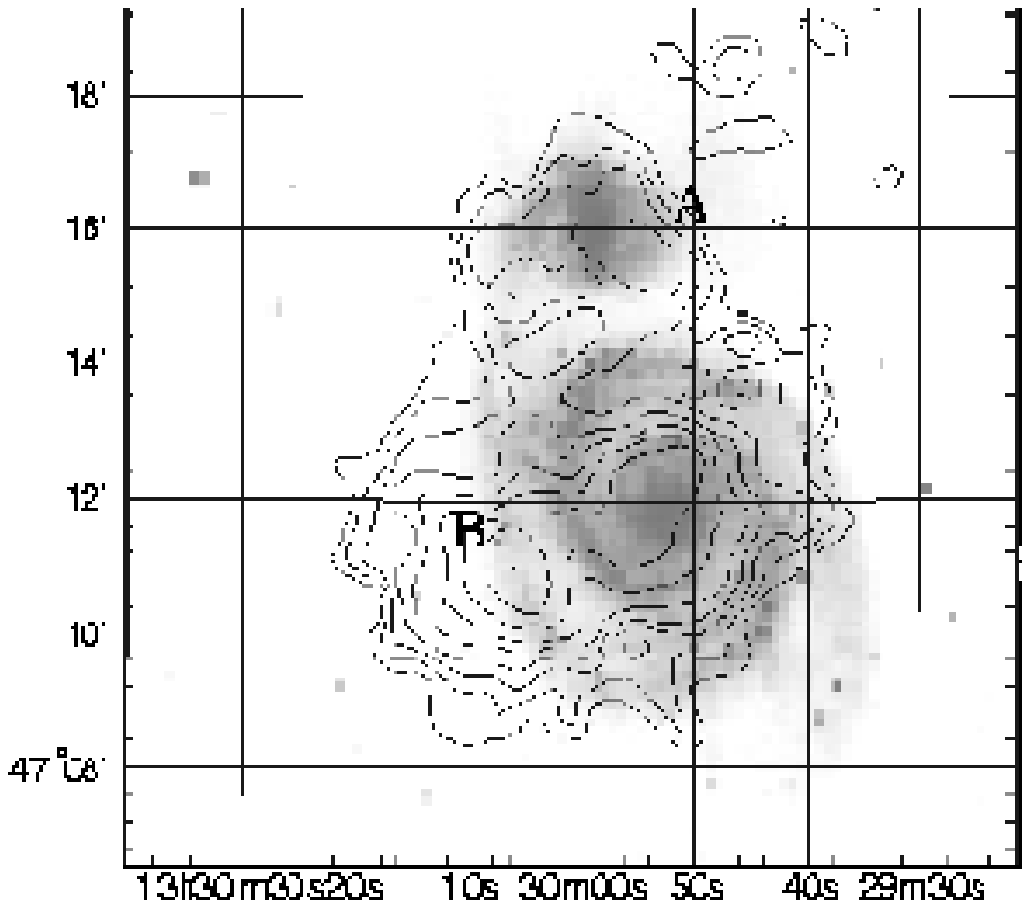}}}
\end{minipage}\quad
\begin{minipage}{8cm}
\centerline{\resizebox{8cm}{!}{\includegraphics{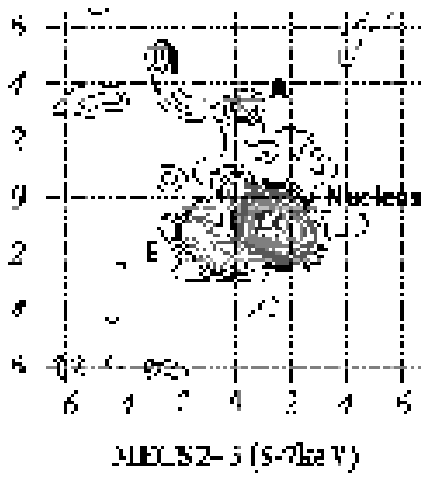}}}
\end{minipage}
\caption{BeppoSAX MECS images of M~51. (a) MECS image
 (contour) in the total band overlaid on the optical image (gray scale).
(b)MECS image in the energy band of 5--7 keV. The center position is (13h
 30m 00s, 47d 13m 00s). The axis unit is arcmin. }
\end{figure}

\begin{figure}
\centerline{\resizebox{15cm}{!}{\includegraphics{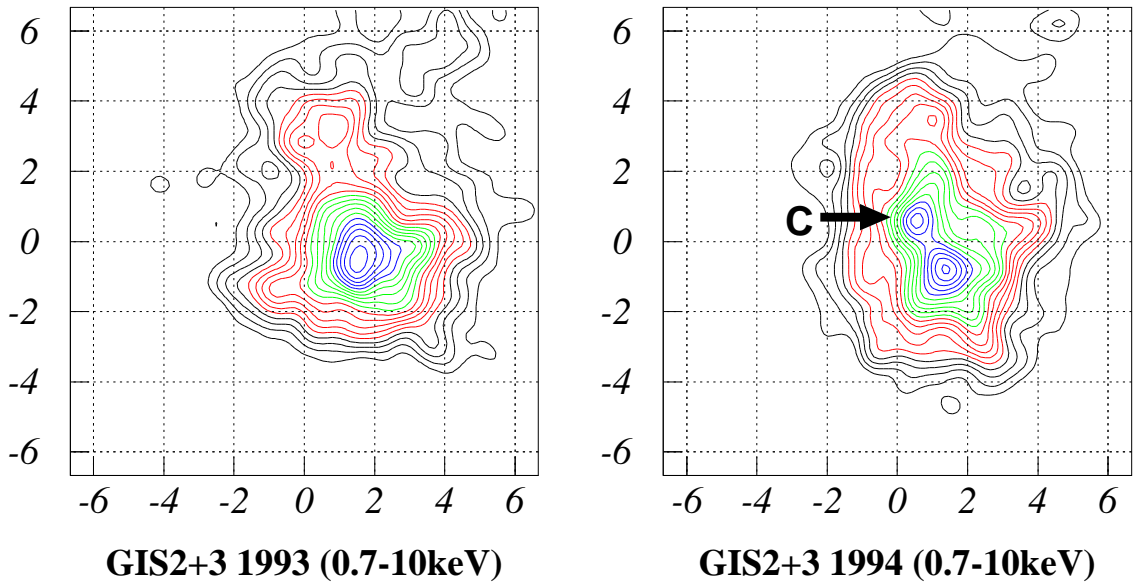}}}
\caption{ASCA GIS images of M~51 in 1993 (left) and 1994 (right). The axis unit is arcmin.}
\end{figure}

\begin{figure}
\centerline{\resizebox{8cm}{!}{\includegraphics{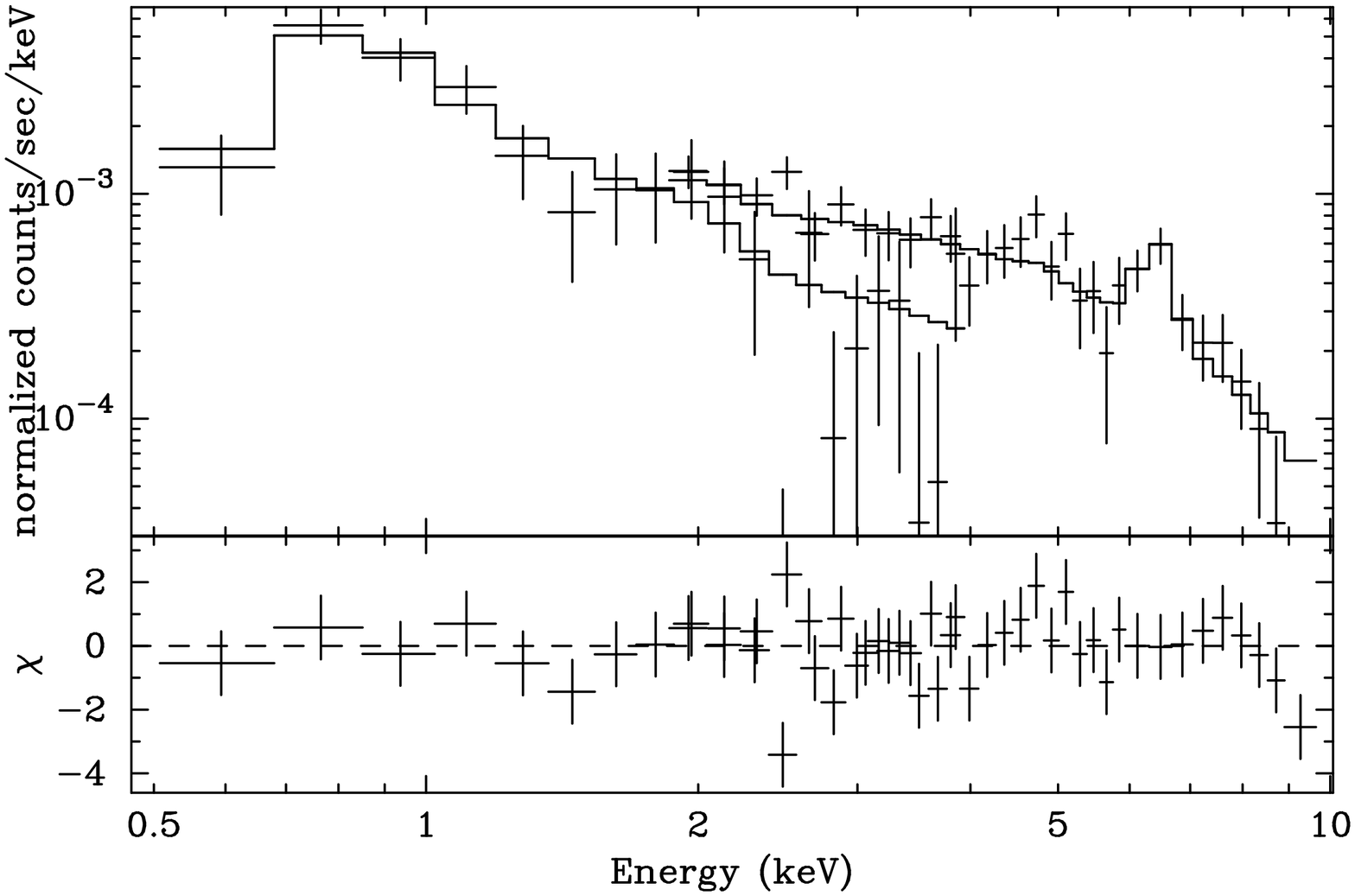}}}
\caption{LECS and MECS spectra of the nucleus. The solid line represents the 
 best-fit model (Raymond-Smith model + power law model + Gaussian).}
\end{figure}

\begin{figure}
\begin{minipage}{8cm}
\centerline{\resizebox{8cm}{!}{\includegraphics{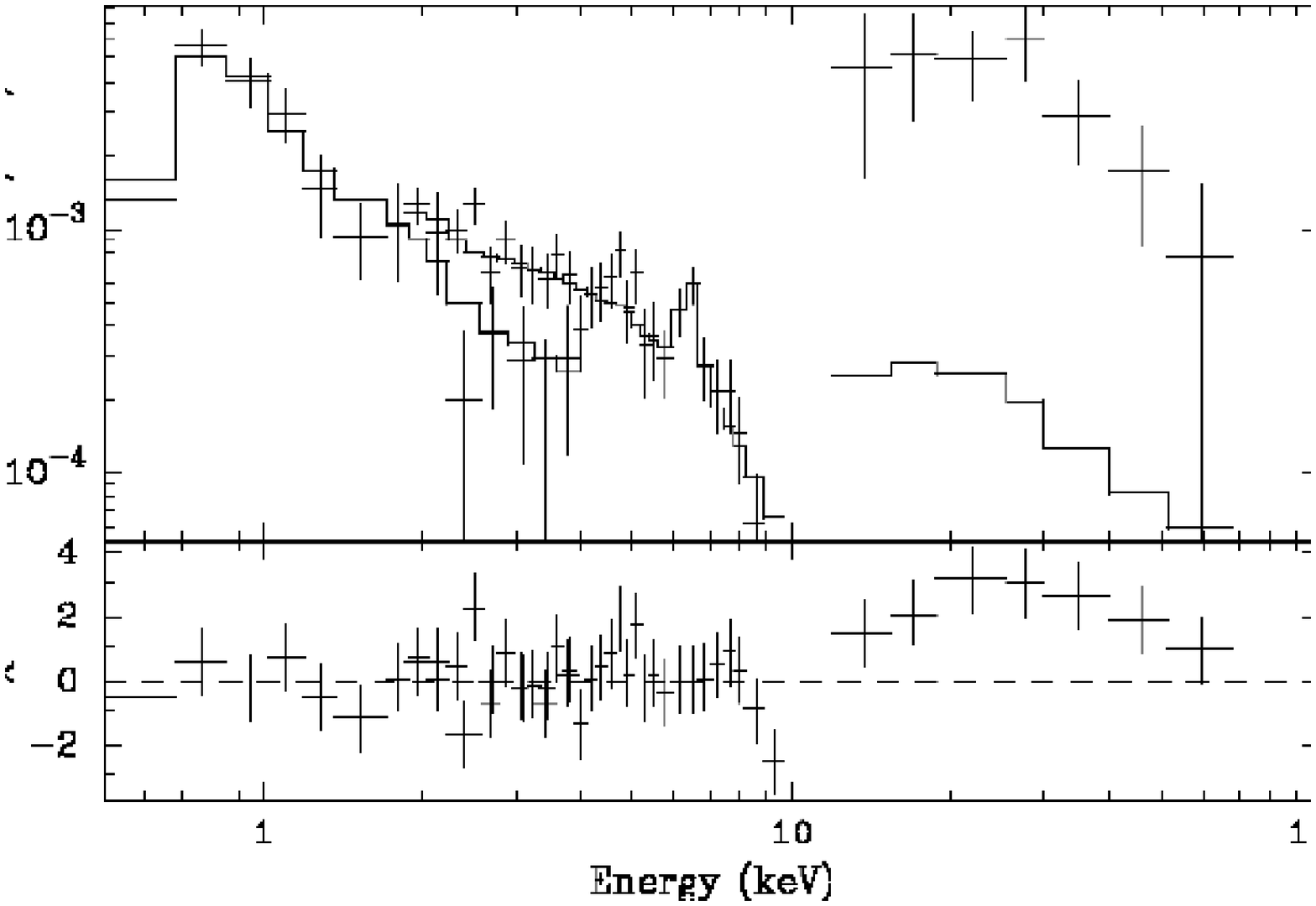}}}
\end{minipage}\quad
\begin{minipage}{8cm}
\centerline{\resizebox{8cm}{!}{\includegraphics{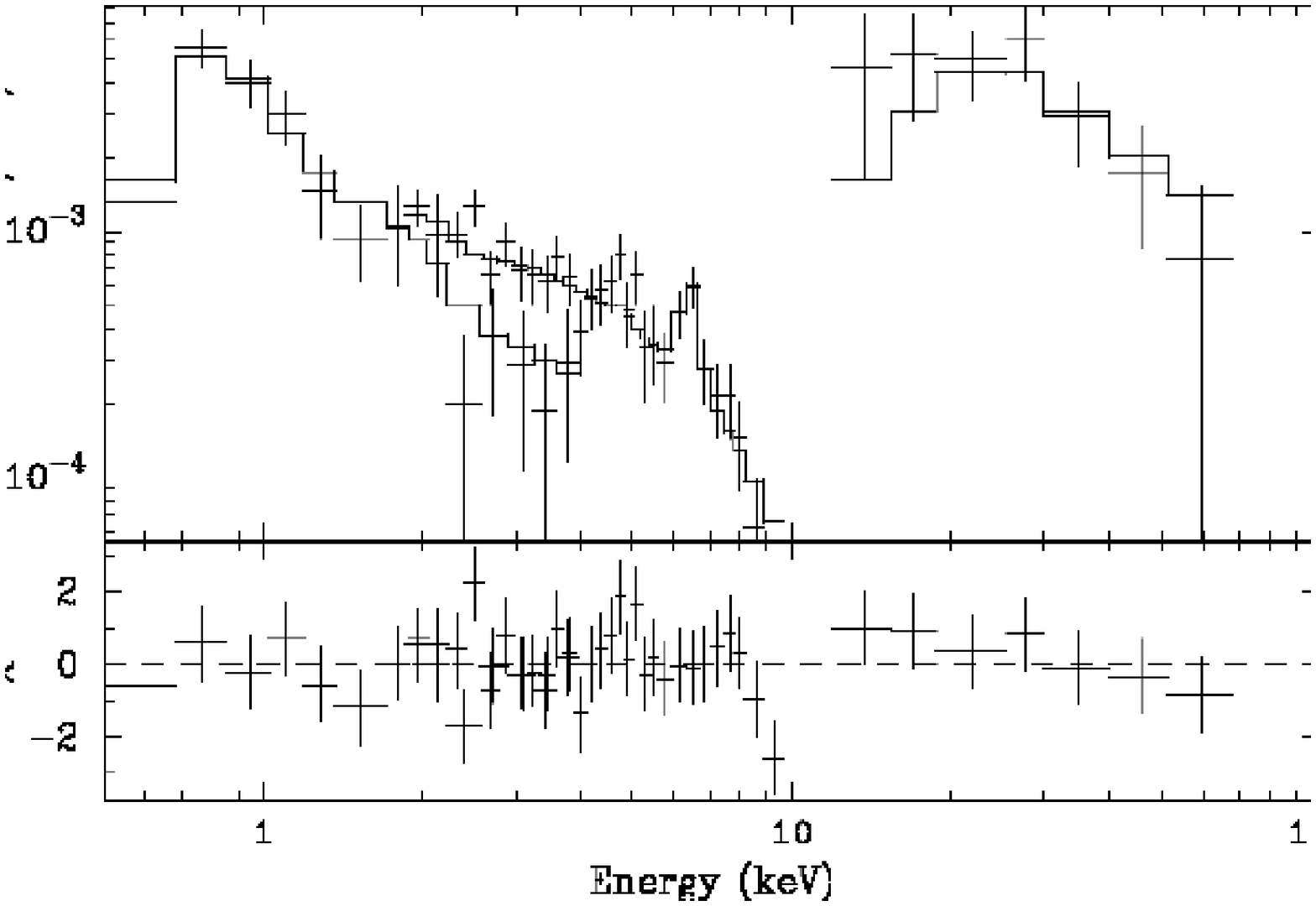}}}
\end{minipage}
\caption{LECS, MECS, and PDS spectra of the nucleus. 
The solid line represents the best-fit model (the same as Fig. 4) for
 the LECS and MECS spectra (left), and the best-fit model (Raymond-Smith 
model + power law model + Gaussian + absorbed power law) for all the
 instruments (right).}
\end{figure}

\begin{figure}
\centerline{\resizebox{8cm}{!}{\includegraphics{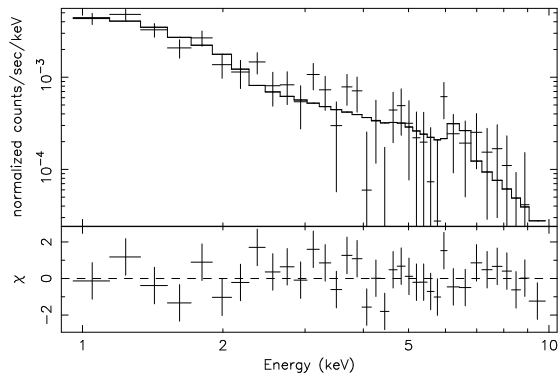}}}
\caption{ASCA GIS spectrum of the nucleus in 1994. The solid line
 represents the best-fit model (see text).}
\end{figure}

\begin{figure}
\centerline{\resizebox{8cm}{!}{\includegraphics{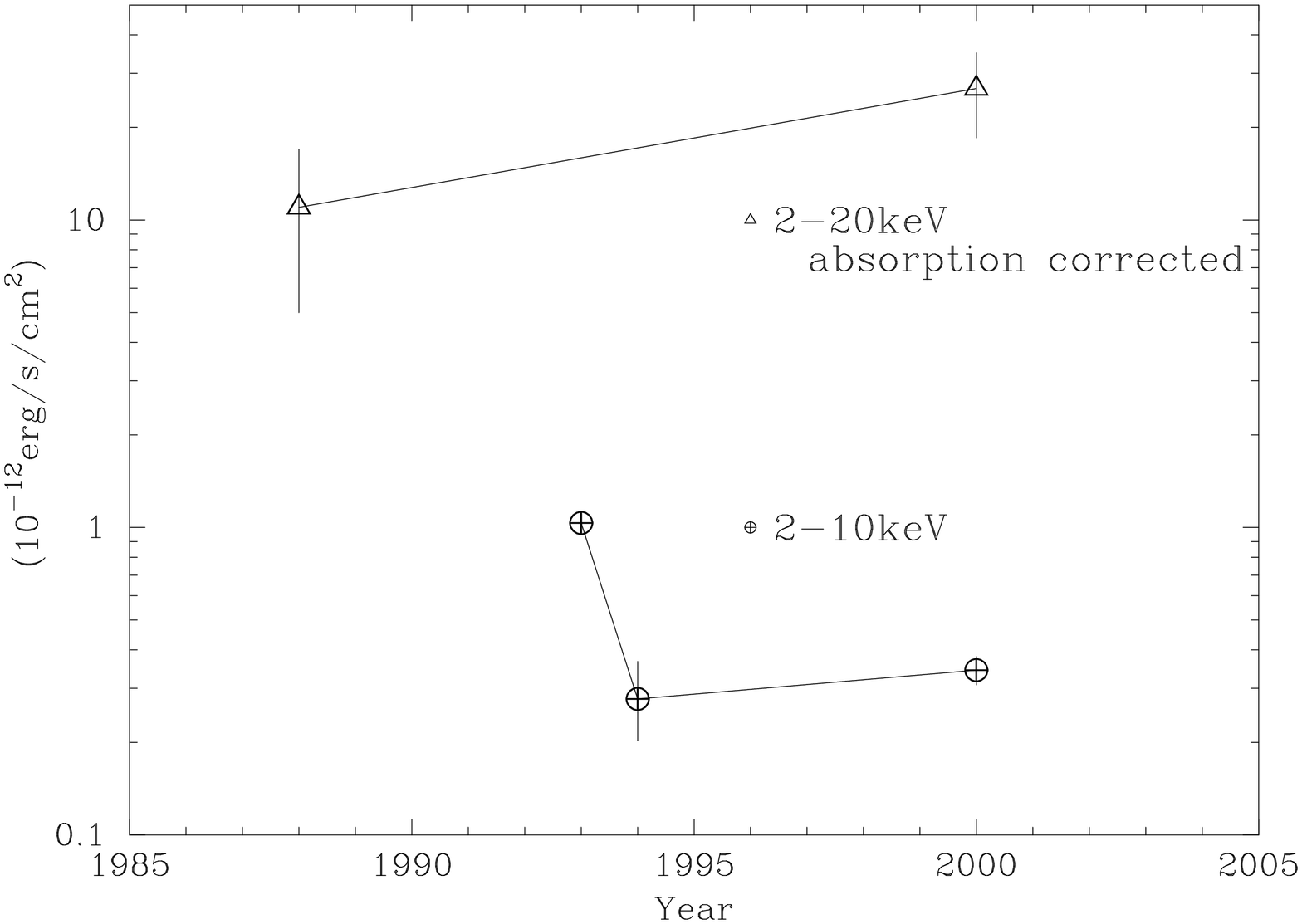}}}
\caption{Light curve of the M~51 nucleus measured 
by various satellites. The open
 circles represent the flux of the weak continuum in 2 -- 10 keV
 band, which are thought to be a scattered or reflected component. 
The open triangles are the flux of the bright nuclear direct 
continuum in 2 -- 20 keV band.}
\end{figure}

\newpage
\begin{figure}
\centerline{\resizebox{8cm}{!}{\includegraphics{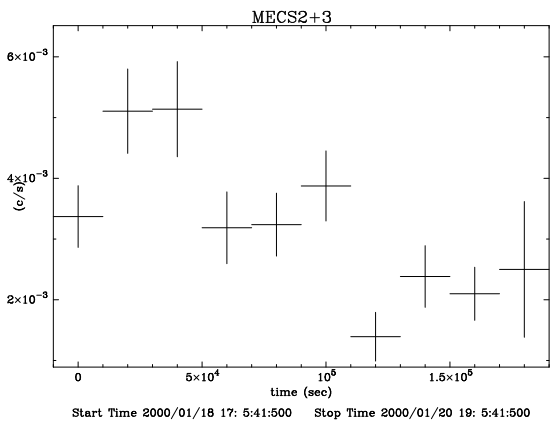}}}
\caption{MECS light curve of the off-nuclear source B.}
\end{figure}

\newpage
\begin{figure}
\centerline{\resizebox{8cm}{!}{\includegraphics{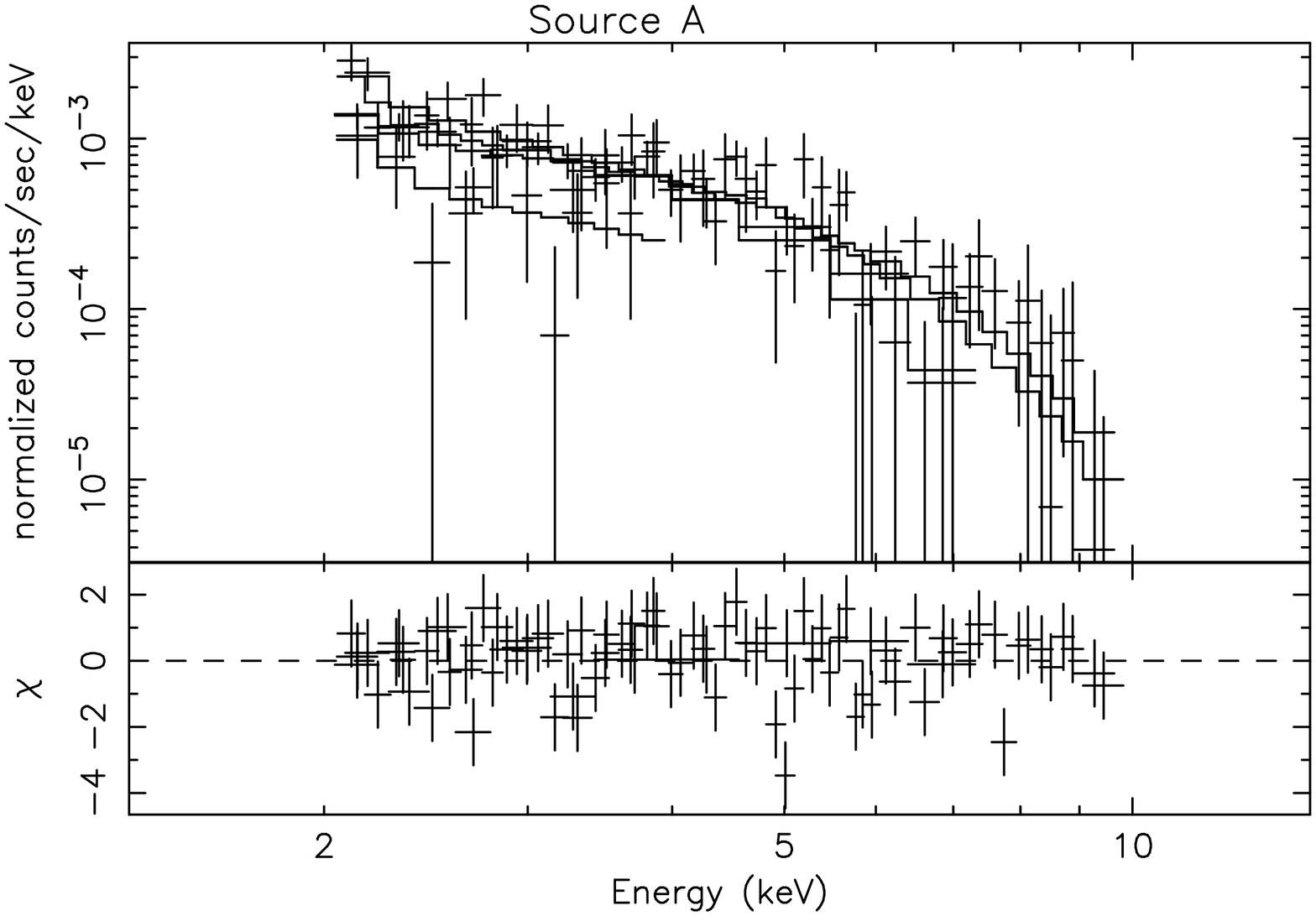}}}
\begin{minipage}{8cm}
\centerline{\resizebox{8cm}{!}{\includegraphics{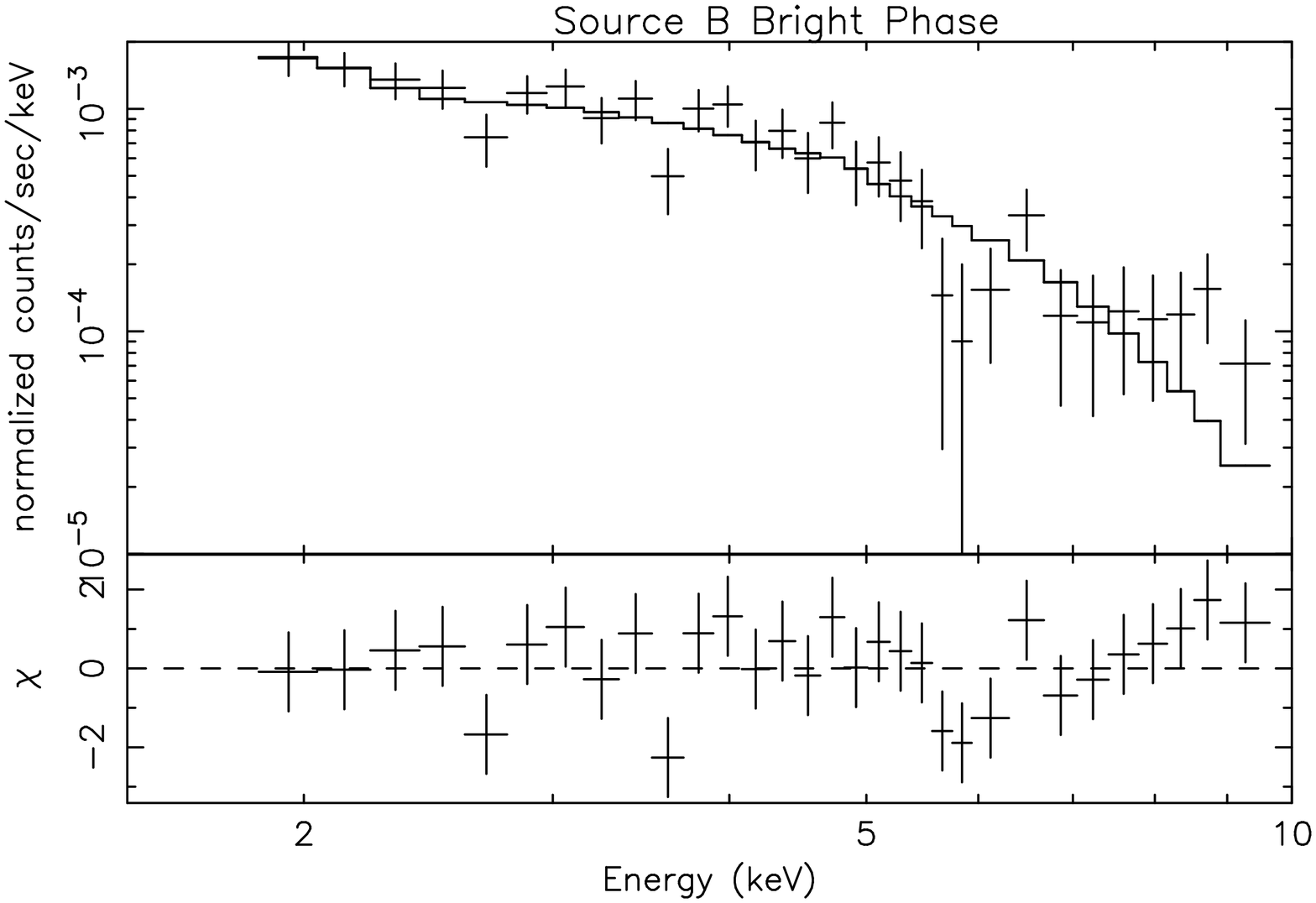}}}
\end{minipage}\quad
\begin{minipage}{8cm}
\centerline{\resizebox{8cm}{!}{\includegraphics{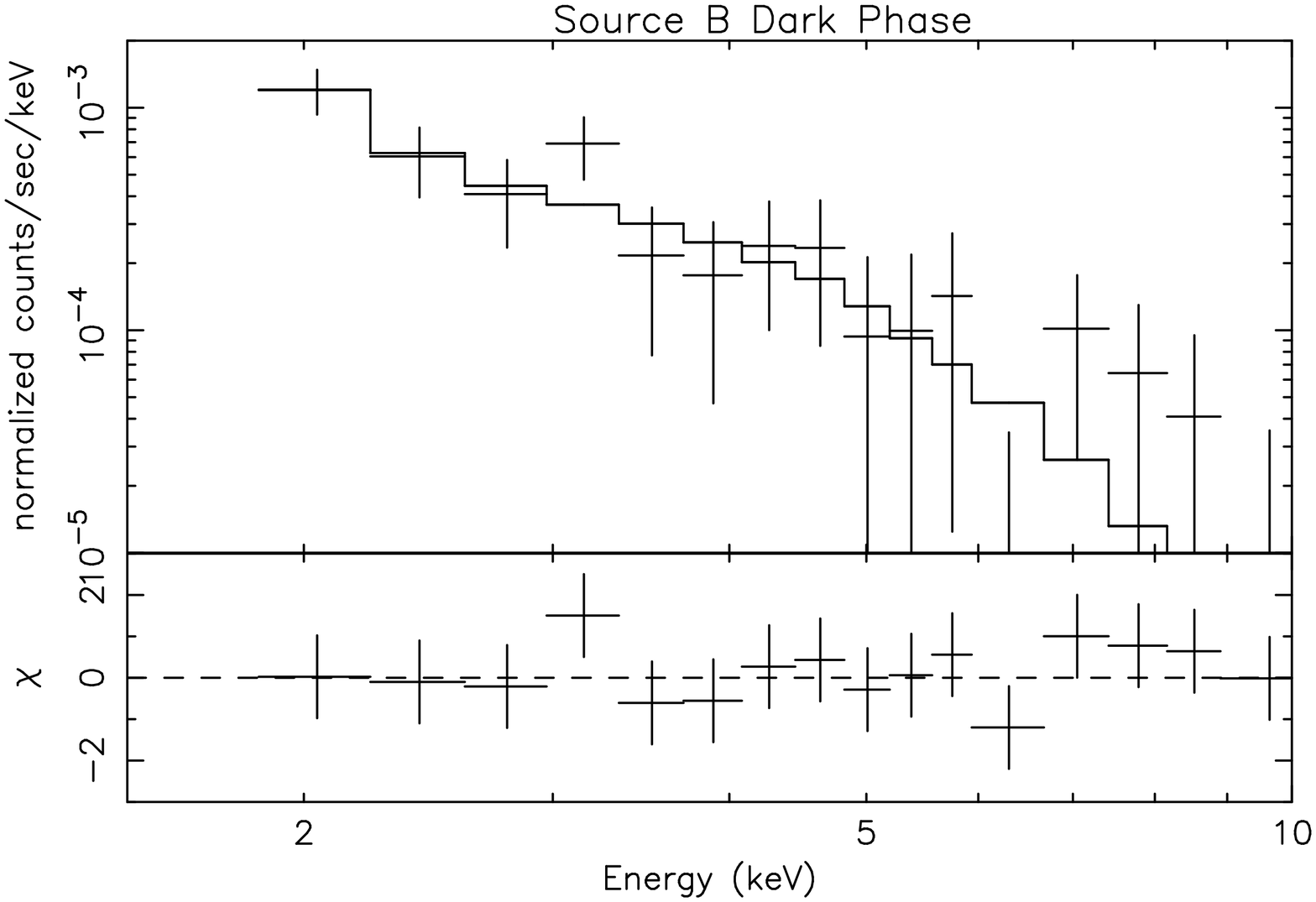}}}
\end{minipage}
\caption{X-ray spectra of the source A and B. (a)ASCA GIS/SIS (1993) 
 and BeppoSAX LECS/MECS (2000) spectra of the source A. (b)MECS
 spectra of the source B. The left and right are obtained in the bright
 and faint phase, respectively.}
\end{figure}

\end{document}